\documentclass[showpacs,10pt,twocolumn,prb]{revtex4-1}
\usepackage{amsmath}
\usepackage{amssymb}
\usepackage{graphics}
\usepackage{epsfig}
\usepackage{color}
\usepackage{lineno}

\begin{document}


\title{Ferromagnetic van der Waals crystal VI$_{3}$}
\author{Shangjie Tian,$^{1,\dag}$ Jian-Feng Zhang,$^{1,\dag}$ Chenghe Li,$^{1}$ Tianping Ying,$^{2}$ Shiyan Li,$^{2,3}$ Xiao Zhang,$^{4,*}$ Kai Liu,$^{1,*}$ and Hechang Lei$^{1,*}$}
\affiliation{$^{1}$Department of Physics and Beijing Key Laboratory of Opto-electronic Functional Materials $\&$ Micro-nano Devices, Renmin University of China, Beijing 100872, China\\
$^{2}$ State Key Laboratory of Surface Physics, Department of Physics, and Laboratory of Advanced Materials, Fudan University, Shanghai 200438, China\\
$^{3}$ Collaborative Innovation Center of Advanced Microstructures, Nanjing 210093, China\\
$^{4}$State Key Laboratory of Information Photonics and Optical Communications $\&$ School of Science, Beijing University of Posts and Telecommunications, Beijing 100876, China}
\date{\today}

\begin{abstract}
We report structural, physical properties and electronic structure of van der Waals (vdW) crystal VI$_{3}$. Detailed analysis reveals that VI$_{3}$ exhibits a structural transition from monoclinic $C2/m$ to rhombohedral $R\bar{3}$ at $T_{s}\sim$ 79 K, similar to CrX$_{3}$ (X = Cl, Br, I). Below $T_{s}$, a long-range ferromagnetic (FM) transition emerges at $T_{c}\sim$ 50 K. The local moment of V in VI$_{3}$ is close to the high-spin state V$^{3+}$ ion ($S=$ 1). Theoretical calculation suggests that VI$_{3}$ may be a Mott insulator with the band gap of about 0.84 eV.
In addition, VI$_{3}$ has a relative small interlayer binding energy and can be exfoliated easily down to few layers experimentally. Therefore, VI$_{3}$ is a candidate of two-dimensional FM semiconductor. It also provides a novel platform to explore 2D magnetism and vdW heterostructures in $S=$ 1 system.
\end{abstract}


\maketitle

\section{Introduction}

Two-dimensional (2D) materials have induced great interest since the discovery of graphene.\cite{Novoselov} They exhibit various exotic physical properties, such as
enhanced direct-gap semiconductor in monolayer MoS$_{2}$,\cite{Mak} Ising superconductivity in ion-gated MoS$_{2}$ and atomic layer NbSe$_{2}$,\cite{LuJM,XiX} robust in-plane
ferroelectricity in atomic-thick SnTe,\cite{KChang} and so on.
In theory, the long-range 2D magnetic order is strongly suppressed above zero temperature for isotropic Heisenberg model because the fluctuation is so strong that will destroy the spontaneous symmetry breaking (Merin-Wagner theorem).\cite{Mermin} But when magnetic anisotropy is included, the 2D ferromagnetism can be stabilized and it has been realized in monolayer Cr$_{2}$Ge$_{2}$Te$_{6}$ and CrI$_{3}$ just in recent.\cite{GongC,HuangB} These 2D ferromagnets provide unique opportunities for understanding, exploring and utilizing novel low-dimensional magnetism.
Because of its thickness of one or few unit cells, it may be able to control the 2D magnetic properties by applying electric field, light or strain. Such magneto-electrical, magneto-optical, or spin-lattice coupling effects are usually difficult to realize in bulk crystals. For instance, recent studies show that the electrostatic doping can control the magnetism in atomically-thin CrI$_{3}$ and Fe$_{3-x}$GeTe$_{2}$ flakes, leads to the dramatic enhancement of $T_{c}$ from about 100 K to room temperature for the latter.\cite{JiangSW,DengYJ} Moreover, the weak van der Waals (vdW) interlayer coupling allows to design heterostructures between these 2D magnets and other vdW materials without considering lattice mismatch. For example, in the vdW heterostructures formed by an ultrathin CrI$_{3}$ and a monolayer WSe$_{2}$, the WSe$_{2}$ photoluminescence intensity strongly depends on the relative alignment between photoexcited spins in WSe$_{2}$ and the CrI$_{3}$ magnetization.\cite{ZhongD}

Besides Cr$_{2}$Ge$_{2}$Te$_{6}$ and CrI$_{3}$, CrSiTe$_{3}$ and CrX$_{3}$ (X = Cl and Br) vdW crystals are also easy to be exfoliated and predicted to be a 2D ferromagnetic (FM) semiconductors for monolayers.\cite{LinMW,ZhangWB,McGuire2} Until now, however, such magnetic vdW materials are still scarce and they are mainly chromium compounds with high spin Cr$^{3+}$ ($S=$ 3/2).
In recent, the 2D ferromagnetism has been predicted in VX$_{3}$ (X = Cl and I) with $S=$ 1.\cite{HeJJ} It is found that VX$_{3}$ is cleavable and the calculated Curie temperature $T_{c}$ is 80 and 98 K for VCl$_{3}$ and VI$_{3}$ monolayers, respectively, higher than those in CrX$_{3}$.

In this work, we grow VI$_{3}$ single crystals successfully and carry out a comprehensive study on their structure, physical properties and electronic structure.
VI$_{3}$ exhibits a structural transition from monoclinic to rhombohedral at $T_{s}\sim$ 79 K and then a long-range FM ordering appears at $T_{c}\sim$ 50 K.
Moreover, VI$_{3}$ should be a Mott insulator and can be exfoliated easily to few layers.

\section{Methods}

\textbf{Single Crystal Growth.} Single crystals of VI$_{3}$ were grown by chemical vapor transport method. Vanadium powder (99.99\%) and iodine flake (99.999\%) in 1 : 3 molar ratio were put into a silica tube with the length of 200 mm and the inner diameter of 14 mm. The tube was evacuated down to 10$^{-2}$ Pa and sealed under vacuum. The tubes were placed in two-zone horizontal tube furnace and the source and growth zones were raised to 923 K and 823 K for 3 days, and then held there for 7 days. The shiny black plate-like crystals with lateral dimensions up to several millimeters can be obtained. Because the VI$_{3}$ is hygroscopic, the samples were stored in glovebox in order to avoid degradation of crystals.

\textbf{X-ray Crystallography.} Single crystal X-ray diffraction (XRD) patterns at 40 K, 60 K and 100 K were collected using a Bruker D8 VENTURE PHOTO II diffractometer with multilayer mirror monochromatized Mo K$\alpha$ ($\lambda=$ 0.71073 \AA) radiation. Unit cell refinement and data merging were done with the SAINT program, and an absorption correction was applied using Multi-Scans. At $T=$ 100 K, an inspection of the systematic absences immediately showed the structure to be $C$ centered. In XPrep, the suggested possible space group is $C2$, $C2/m$ and $Cm$ with the value of combined figure of merit (CFOM) 11.86, 2.01 and 14.71, respectively. The mean value of $\mid E^{2}-1\mid$ (1.067) is more close to the value for a centrosymmetric space group (0.968) than that for a non-centrosymmetric one (0.736), suggesting the centrosymmetric structure of VI$_{3}$. Finally the space group $C2/m$ with the smallest CFOM value is selected. When $T=$ 40 K, an inspection of the systematic absences immediately showed the structure to be $R$ lattice. The possible space group is $R\bar{3}$, $R3$, $R3m$, $R32$ and $R\bar{3}m$ with the vaule of CFOM 7.68, 25.36, 47.09, 47.09 and 31.27, respectively. The mean value of $\mid E^{2}-1\mid$ (1.213) also suggested that the space group is centrosymmetric, thus the space group $R\bar{3}$ with the smallest CFOM value is selected. A structural solutions with the $C2/m$ and $R\bar{3}$ space group were obtained for VI$_{3}$ by intrinsic phasing methods using the program APEX3,\cite{APEX} and the final refinement was completed with the SHELXL suite of programs.\cite{Sheldrick}

\textbf{Measurements of Physical Properties.} The magnetic properties were measured using a Quantum Design magnetic property measurement system (MPMS-3). A piece of VI$_{3}$ single crystal was mounted in the plastic straw with different orientations in order to measure the anisotropy of magnetization. Heat capacity measurements were performed using a Quantum Design physical property measurement system (PPMS-14T). Exfoliation of bulk VI$_{3}$ was achieved using mechanical exfoliation with scotch tape and transferred onto a 90 nm SiO$_{2}$ covered Si substrate. Atomic force microscopy was carried out using a Bruker Edge Dimension atomic force microscope.

\textbf{Theoretical Calculation.} The fully spin-polarized electronic structure calculations were carried out with the projector augmented wave (PAW) method \cite{PAW} as implemented in the VASP package.\cite{VASP1,VASP2} The generalized gradient approximation (GGA) of Perdew-Burke-Ernzerhof (PBE) type \cite{PBE} was chosen for the exchange-correlation functional. The kinetic energy cutoff of the plane-wave basis was set to be 400 eV. The low-temperature (40 K) $R\bar{3}$ crystal structure with the experimental lattice constants (Table I) was adopted. Four typical intralayer magnetic configurations,\cite{HeJJ} i.e., the ferromagnetic (FM) state, the checkerboard antiferromagnetic (AFM) N\'eel state, the AFM stripy state, and the AFM zigzag state, were investigated. These magnetic configurations were studied with a $\sqrt{3}\times1\times1$ orthorhombic supercell, which contains four V atoms in each layer, and a $4\times8\times4$ $k$-point mesh. The Fermi level was broadened by the Gaussian smearing method with a width of 0.05 eV. The electronic correlation effect among V 3$d$ electrons was incorporated by using the GGA + U formalism of Dudarev et al..\cite{DFTU} with the effective Hubbard $U$ of 3.68 eV.\cite{HeJJ} The vdW interactions between the VI$_{3}$ layers were considered by adopting the optB86b-vdW functional.\cite{vdW} The internal atomic positions were allowed to relax until all forces on atoms were smaller than 0.01 eV/\AA. The spin-orbit coupling (SOC) effect was included when the equilibrium atomic positions were obtained. The binding energy $E_{\rm B}$ was calculated by increasing the interlayer distance $d$ of a $R\bar{3}$ conventional cell.\cite{Eb}

\section{Results and Discussion}

\begin{figure}[tbp]
\centerline{\includegraphics[scale=0.25]{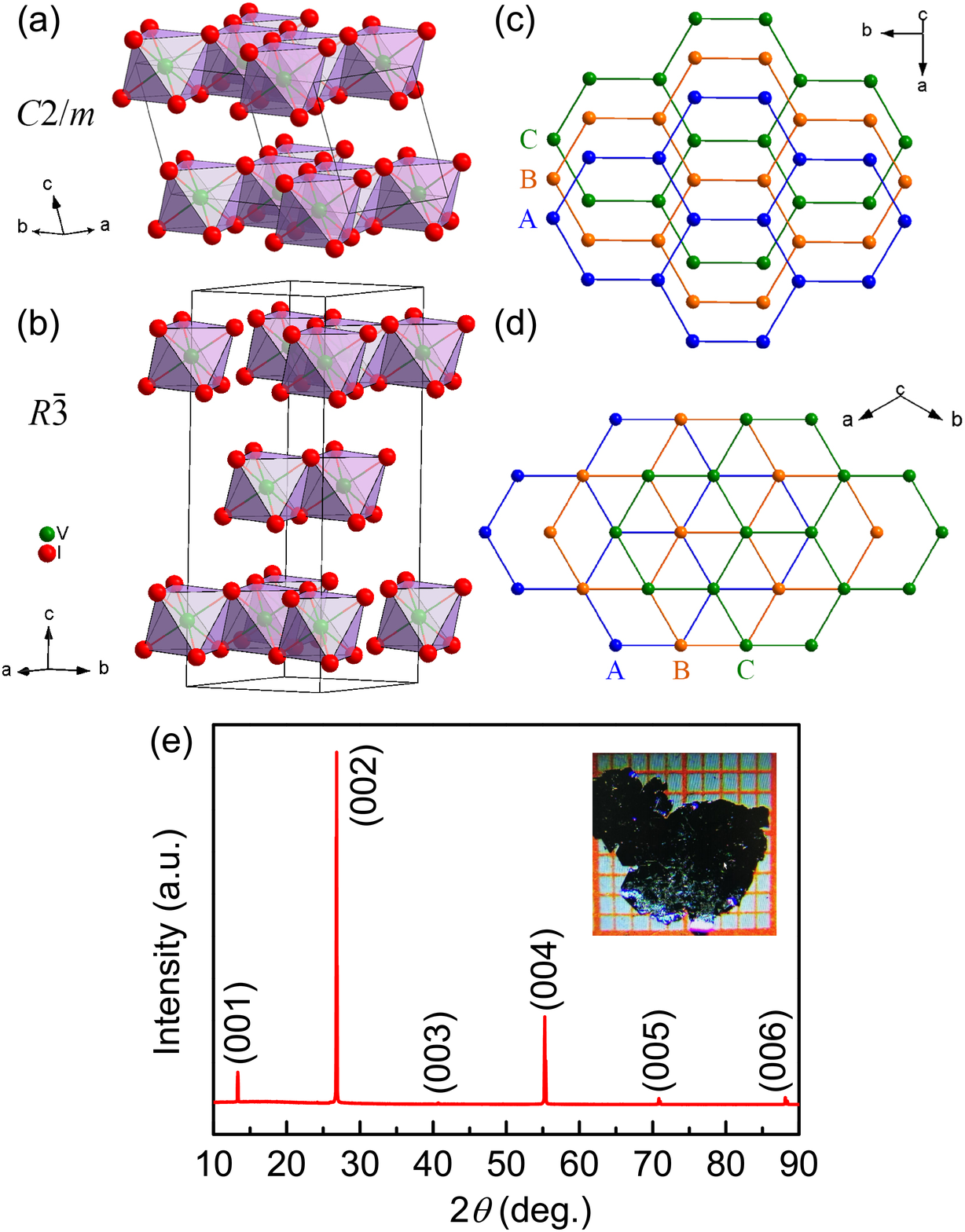}} \vspace*{-0.3cm}
\caption{(a) and (b) High- and low-temperature crystal structures of VI$_{3}$, respectively. (c) and (d) Stacking pattern of the V layers in the high- and low-temperature structures of VI$_{3}$, respectively. The labels of A, B and C denote the different layers of V atoms. (e) XRD pattern of a VI$_{3}$ single crystal. Inset: photo of typical VI$_{3}$ single crystals on a 1 mm grid paper.}
\end{figure}

\begin{table*}
\centering
\caption{Crystallographic Data for VI$_{3}$ at Different Temperatures.}
\tabcolsep 0.2in
\renewcommand\arraystretch{1.5}
\begin{tabular}{cccc}
\hline
$T$ (K) & 40 & 60 & 100   \\
space group & $R\bar{3}$ & $R\bar{3}$ & $C2/m$  \\
crystal system & trigonal & trigonal & monoclinic  \\
$a$ ({\AA}) & 6.8351(7) & 6.8325(6) & 6.8416(3) \\
$b$ ({\AA}) & 6.8351(7) & 6.8325(6) & 11.8387(6) \\
$c$ ({\AA}) & 19.696(2) & 19.6776(2) & 6.9502(4) \\
$V$ ({\AA}$^{3}$) & 796.90(19) & 795.54(16) & 533.66(36) \\
$Z$    & 6 & 6 & 4 \\
dimens min/mid/max(mm$^{3}$) & 0.03/0.28/0.37 & 0.03/0.28/0.37 & 0.03/0.28/0.37 \\
calcd density (g cm$^{-3}$) & 5.397 & 5.406 & 5.372 \\
abs coeff (mm$^{-1}$)& 19.117 & 19.149 & 19.031 \\
$h$ & -9 $\leq$ $h$ $\leq$ 9 & -8 $\leq$ $h$ $\leq$ 8 & -9 $\leq$ $h$ $\leq$ 9  \\
$k$ &  -8 $\leq$ $k$ $\leq$ 7 & -9 $\leq$ $k$ $\leq$ 9 & -15 $\leq$ $k$ $\leq$ 15  \\
$l$ &  -26 $\leq$ $l$ $\leq$ 26 & -26 x$\leq$ $l$ $\leq$ 26 & -9 $\leq$ $l$ $\leq$ 9  \\
reflns collected/unique/$R$(int)& 3799/442/0.0409 & 4793/442/0.0471 & 4901/681/0.0417  \\
data/params/restraints  & 442/15/0 & 442/15/0 & 681/22/0  \\
GOF on $F^{2}$ & 1.228 & 1.287 & 1.025  \\
$R$ indices (all data) ($R$1/$wR$2)$^{a}$ & 0.0636/0.1691 & 0.0502/0.1549 & 0.0724/0.1828  \\
\hline
\end{tabular}
\end{table*}

\begin{table}
  \centering
\caption{Atomic Positions and Equivalent Isotropic Displacement Parameters $U_{\rm eq}$ for VI$_{3}$ at Different Temperatures.}
\begin{tabular}{cccccc}
\hline
atom & site & $x/a$ & $y/b$ & $z/c$ & $U_{\rm eq}$ (A$^{2}$)  \\
40 K &&&&&\\
V & 6$c$ & 1/3 & 2/3 & 0.50035(11) & 0.0056(7)  \\
I & 18$f$ & 0.00134(7) & 0.6509(2) & 0.42086(3) & 0.0074(5)  \\
60 K &&&&&\\
V & 6$c$ & 1/3 & 2/3 & 0.50035(11) & 0.0052(7)  \\
I & 18$f$ & 0.00143(8) & 0.6507(2) & 0.42086(3) & 0.0069(5)  \\
100 K &&&&&\\
V & 4$g$ & 0 & 0.16658(18) & 0 & 0.0136(7)  \\
I1& 4$i$ & 0.2255(2) & 0 & 0.23895(15) & 0.0167(5)  \\
I2& 8$j$ & 0.25103(10) & 0.32498(8) & 0.23547(13) & 0.0163(5)  \\
\hline
\end{tabular}
\end{table}

As shown in Table I and II, the crystal structure of VI$_{3}$ at high temperature is not rhombohedral BiI$_{3}$ type reported in previous study on the VI$_{3}$ polycrystal,\cite{Berry} but monoclinic AlCl$_{3}$ type (space group $C2/m$, No. 12), similar to high-temperature structure of CrX$_{3}$.\cite{Morosin,McGuire,McGuire2}
For this structure (Fig. 1(a)), the V$^{3+}$ cations are surrounded by six I$^{-}$ anions with an octahedral coordination. These edge-shared distorted octahedra form honeycomb layers in the $ab$ plane. The slabs of VI$_{3}$ are stacked along the $c$ axis and there are vdW gaps between them.
On the other hand, when decreasing temperature, there is a structural transition from monoclinic phase to rhombohedral one with space group $R\bar{3}$ (Tables I and II, Fig. 1(b)), same as CrX$_{3}$ (X = Cl, Br, I).\cite{Morosin,McGuire,McGuire2}
The key difference between these two structures is the stacking patterns of V-I slabs along the $c$ axis. For the monoclinic structure, the V-I slab shifts layer-by-layer along the $a$ axis. Because the $\beta$ is about 108.558$^{\circ}$ and the $a$ and $c$ axial lattice parameters are 6.8416 and 6.9502 \AA, the $a/[c\cos(180^{\circ}-\beta)]$ is 3.09, i.e., the stacking is nearly the ABC fashion (Fig. 1(c)). For the rhombohedral structure (Fig. 1(b)), the stacking of V-I slabs along the $c$ axis is exactly the ABC fashion, i.e., each V-I layer shifts the distance of V-V bond ($\sim$ 3.95 \AA) along one edge of V honeycomb (the [$\bar{1}$10] direction) (Fig. 1(d)). This shift direction is distinctly different from that perpendicular to one edge of V honeycomb in the monoclinic structure.
In contrast to the drastic change of stacking order, the intralayer structure is almost intact, except a slight decreased distortion of V honeycomb and V-I octahedron in rhombohedral structure. The bond lengths of V-I are spread between 2.7122 and 2.7195 \AA\ in monoclinic structure and narrow down to 2.7173 and 2.7179 \AA\ for the low-temperature phase (Table III). Moreover, in high-temperature structure, the V honeycomb is distorted with two bond lengths of V-V (3.9442 and 3.9501 \AA). It becomes undistorted in low-temperature structure and the single bond length of V-V is about 3.9463 \AA.
In the XRD pattern of a VI$_{3}$ single crystal (Fig. 1(e)), only the $(00l)$ reflections of monoclinic structure can be observed, indicating that the surface of crystal is parallel to the $ab$ plane. The single crystals of VI$_{3}$ show plate-like shapes (inset of Fig. 1(e)), consistent with the single crystal XRD pattern and the layered structure of VI$_{3}$.
The actual atomic ratio of V : I determined from the EDX analysis is 1.00 : 2.90(1) when setting the content of V as 1, in agreement with the nominal formula of VI$_{3}$.

\begin{table}
\centering
\caption{Selected Bond Lengths (in \AA) for VI$_{3}$ at Different Temperatures.}
\tabcolsep 0.2in
\renewcommand\arraystretch{1.5}
\begin{tabular}{ccc}
\hline
40 K &&\\
V-I & 3$\times$ & 2.7143(14)  \\
V-I & 3$\times$ & 2.7179(14)  \\
60 K &&\\
V-I & 3$\times$ & 2.7118(15)  \\
V-I & 3$\times$ & 2.7161(14)  \\
100 K &&\\
V-I2 & 2$\times$ & 2.7122(17)  \\
V-I2 & 2$\times$ & 2.7138(8)  \\
V-I1 & 2$\times$ & 2.7195(18)  \\
\hline
\end{tabular}
\end{table}

\begin{figure}[tbp]
\centerline{\includegraphics[scale=0.17]{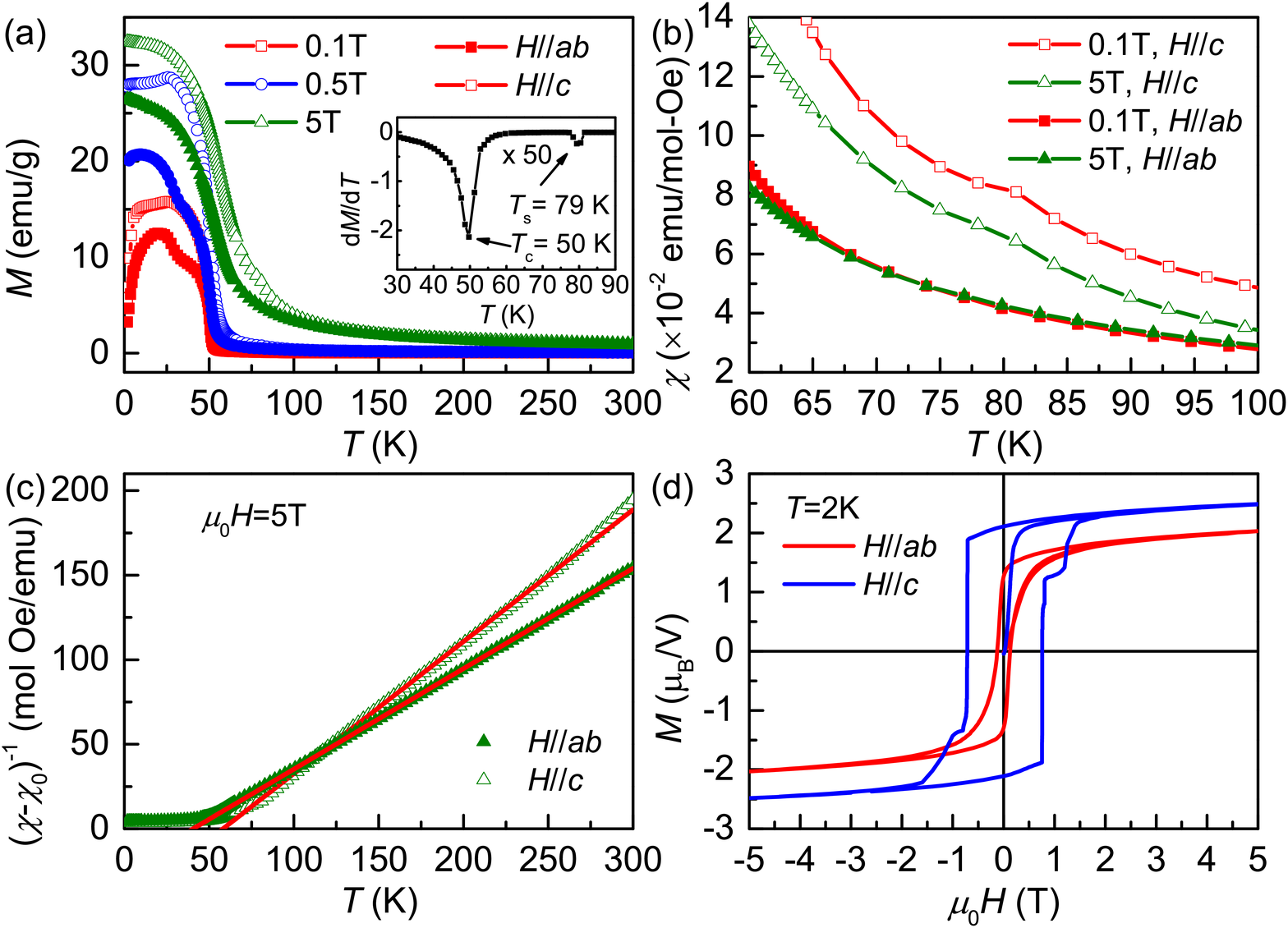}} \vspace*{-0.3cm}
\caption{(a) Temperature dependence of magnetic moment $M(T)$ of VI$_{3}$ single crystal with zero-field-cooling (ZFC) mode at $\mu_{0}H=$ 0.1, 0.5, and 5 T for $H\Vert ab$ and $H\Vert c$. Inset: d$M(T)/$d$T$ vs. $T$ at $\mu_{0}H=$ 0.1 T for $H\Vert c$. (b) enlarged magnetic susceptibility $\chi(T)$ between 60 - 100 K at $\mu_{0}H=$ 0.1 and 5 T for both field directions. (c) Inverse magnetic susceptibility (points) and Curie-Weiss fits (solid lines) as a function of temperature at $\mu_{0}H=$ 5 T for $H\Vert ab$ and $H\Vert c$. Fitted temperature-independent terms $\chi_{0}$ are subtracted. (d) Isothermal $M(H)$ curves at 2 K for both field directions.
}
\end{figure}

Figure 2(a) shows the temperature dependence of magnetic moment $M(T)$ of the VI$_{3}$ single crystal at various fields in the $ab$ plane and along the $c$ axis. It can be clearly seen that VI$_{3}$ exhibits a ferromagnetic transition at low temperature for both field directions. The Curie temperature $T_{c}$ determined from the peak of d$M$/d$T$ at $\mu_{0}H=$ 0.1 T for $H\Vert c$ is 50 K (inset of Fig. 2(a)), consistent with previous result.\cite{Wilson}
Another important feature is a kink appearing on the $\chi_{c}(T)$ curve at $T_{s}\sim$ 80 K ($>T_{c}$) and it can be seen more clearly from the small peak of d$M$/d$T$ at 79 K (inset of Fig. 2(a)). This kink does not shift with the variation of field (Fig. 2(b)). This temperature is in the temperature region of structural transition, thus the anomaly of $\chi_{c}(T)$ curve should originate from the change of crystal structure. In contrast, the $\chi_{ab}(T)$ curves do not show this behavior. It suggests that there is a more stronger coupling of the crystal structure to the magnetism along the stacking direction in VI$_{3}$.
Moreover, from the structure data shown in Table I, there is also remarkable change of the interlay spacing from 6.5888 \AA [=$c\sin(180^{\circ}-\beta)$] at 100 K to 6.5592 \AA [=$c$/3] at 60 K. There is about 0.45 \% change when compared to the variation of 0.13 \% for the intralayer $a$-axial lattice parameters. Such huge changes of interlayer spacing could not be solely explained by the effect of conventional thermal contraction and the structural transition should take significant effect on it.
It is worthy to mention that such subtle change accompanying by the sudden drop interlayer distance when crossing structural phase transition has also been observed in CrX$_{3}$,\cite{McGuire,McGuire2} indicating the spin-lattice coupling might be a common feature in layered magnetic halide compounds.

For both field directions, the $M(T)$ curves between 100 K and 300 K can be fitted very well using the modified Curie-Weiss law $\chi(T)=\chi_{0}+C/(T-\theta)$, where $\chi_{0}$ is a temperature-independent term including core diamagnetism, van Vleck paramagnetism as well as background signal of sample holder, $C$ is the Curie constant and $\theta$ is the Weiss temperature (Fig. 2(c)).
The fitted $C$ and $\theta$ are 1.684(4) [1.281(8)] emu mol$^{-1}$ Oe$^{-1}$ K and 40.5(1) [58.1(3)] K for $H\Vert ab$ [$H\Vert c$]. The fitted positive values of $\theta$ for both field direction are close to the $T_{c}$, confirming the ferromagnetic interaction in VI$_{3}$. The obtained values of $C$ correspond to an effective moment of $\mu_{\rm eff}$ = 3.67(1) and 3.20(2) $\mu_{\rm B}/\rm V$ for $H\Vert ab$ and $H\Vert c$, respectively. These values are somewhat larger than that of spin-only V$^{3+}$ ions (2.83 $\mu_{\rm B}/\rm V$). This suggests that the orbital moment of V$^{3+}$ ions in VI$_{3}$ might have some contribution on effective moment.
On the other hand, when $T<T_{c}$, there is a kink locating at about 30 K, and then a decrease of $M(T)$ at low temperature, which become invisible at high field. Such behavior is not clear at present and maybe originates from a possible magnetic glassy state at low temperature and low field region.
Fig. 2(d) shows the magnetization loops for both field directions at 2 K. The hysteresis loops for both field directions support the FM behavior when $T<T_{c}$.
There is an obvious magnetic anisotropy at low field region. The initial magnetization curve becomes more quickly saturated for $H\Vert c$ than $H\Vert ab$, suggesting the easy axis is the $c$ axis. The coercive field for $H\Vert c$ is much larger than that for $H\Vert ab$ and it also indicates that the easy axis is along the $c$ axis.
For $H\Vert ab$, the $M$ at 5 T is about 2.02 $\mu_{\rm B}$/V, which is close to the expected saturated moment of $gS=$ 2 $\mu_{\rm B}$/V for high-spin state V$^{3+}$ ion. But the $M$(5T) for $H\Vert c$ is about 2.47 $\mu_{\rm B}$/V, slightly larger than the expected value. It may be due to the anisotropy $g$ factor with unquenched orbital angular moment.

\begin{figure}[tbp]
\centerline{\includegraphics[scale=0.16]{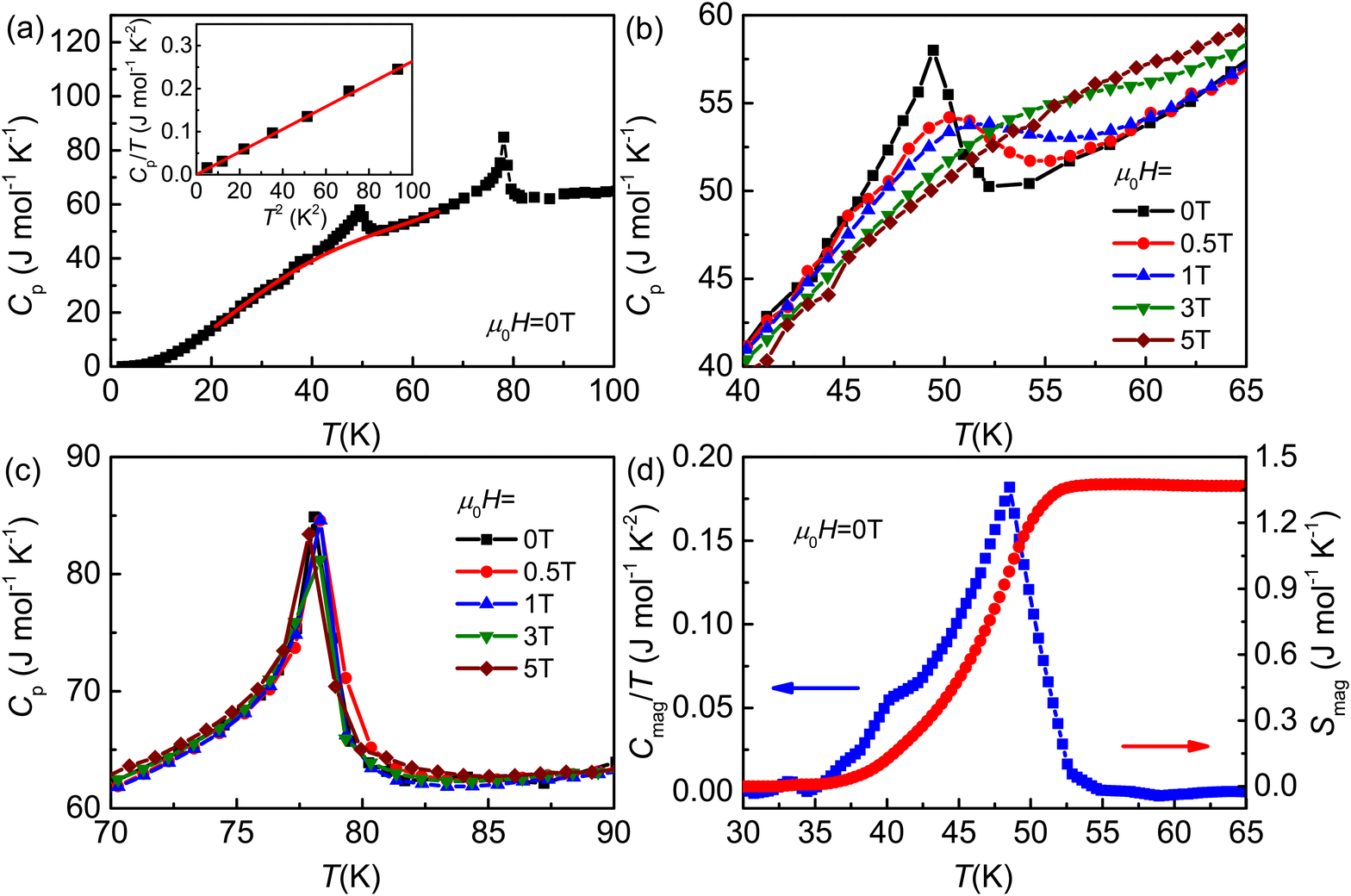}} \vspace*{-0.3cm}
\caption{(a) Temperature dependence of heat capacity $C_{p}(T)$ for VI$_{3}$ single crystal at zero field. The red solid curve represents the lattice contribution, fitted by a polynomial. Inset: enlarged part at low temperature region. The red solid line represents the fit using the formula $C_{p}(T)/T=\beta T^{2}$. (b) The $C_{p}(T)$ as a function of $T$ near magnetic transition region at various fields. (c) Enlarged part of $C_{p}(T)$ near structural transition region at different fields. (d) Temperature dependence of magnetic specific heat $C_{\rm mag}(T)$ after subtracting the lattice contribution fitted by a polynomial. The right axis and its associated solid curve denote the magnetic entropy $S_{\rm mag}(T)$ associated with the FM transition.}
\end{figure}

Figure 3 shows the temperature dependence of heat capacity $C_{p}(T)$ for VI$_{3}$ single crystals measured between $T=$ 2.2 and 100 K at zero field. Clearly, there are two anomalies shown at $T\sim$ 50 K and 79 K, in agreement with those temperatures in magnetization curve for $H\Vert c$. When increasing magnetic field, the $\lambda$-type peak at 49.5 K becomes broaden towards higher temperatures (Fig. 3(b)). This is the typical behavior of FM transition under field, and similar behaviors have been observed in CrX$_{3}$.\cite{McGuire,McGuire2} Thus, it confirms the bulk nature of the long-range FM order below $T_{c}$ observed in the magnetization measurement.
In contrast, the sharp peak at higher temperature ($\sim$ 79 K) is insensitive to external field and there is almost no shift when the field is up to 5 T (Fig. 3(c)). It undoubtedly indicates that this anomaly of heat capacity is corresponding to the structural transition, consistent with the structural characterization.
After subtraction of the phonon contribution $C_{\rm ph}(T)$ fitted using a polynomial from the total heat capacity (red line in the main panel of Fig. 3(a)), we obtain magnetic specific heat $C_{\rm mag}(T)$ and calculate the magnetic entropy $S_{\rm mag}(T)$ using the formula $S_{\rm mag}(T)=\int_{0}^{T}C_{\rm mag}(T)/TdT$ (Fig. 3(d)). The derived $S_{\rm mag}$ is about 1.37 J mol$^{1}$ K$^{-1}$ at 65 K, which is much smaller than the expected value ($\sim $ 15\% R$\ln (2S+1)=$ R$\ln 3$) for V$^{3+}$ ions with high spin state ($S=$ 1).
One of possibilities leading to this difference is the overestimation of $C_{\rm ph}(T)$, reducing the $S_{\rm mag}(T)$. Another possibility is that there is substantial fraction of magnetic entropy released above $T_{c}$ because of possible short-range or 2D magnetic correlations before long-range order forms. But it has to be mentioned that the Weiss temperature $\theta$ determined from the Curie-Weiss law is close to $T_{c}$, implying the magnetic correlations/fluctuation should not so strong when $T>T_{c}$. Similar behaviors are also observed in CrX$_{3}$.\cite{McGuire,McGuire2}, suggesting certain universal features in these vdW magnetic materials. Further theoretical and experimental studies are needed in order to understand these phenomena and the relationship between magnetism and low-dimensional structure.
At the low temperature, the $C_{p}(T)$ curve can be fitted solely by a cubic term $\beta T^{3}$ (inset of Fig. 3(a)) because of the insulating feature of VI$_{3}$ (the resistance is about 1 M$\Omega$ measured using ohmmeter at room temperature).
the fitted value of $\beta $ = 2.63(3) mJ mol$^{-1}$ K$^{-4}$, the Debye temperature is estimated to be $\Theta_{D} $ = 144(1) K using the formula $\Theta _{D}=(12\pi ^{4}NR/5\beta)^{1/3}$. This is slightly larger than that of CrI$_{3}$ ($\Theta_{D}=$ 134 K),\cite{McGuire} possibly ascribed to the smaller atomic mass of V than Cr.

\begin{figure}[tbp]
\centerline{\includegraphics[scale=0.35]{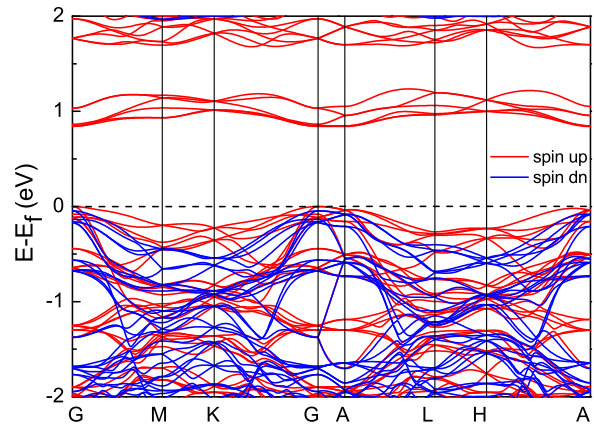}} \vspace*{-0.3cm}
\caption{Calculated electronic band structure of bulk VI$_{3}$ in the ferromagnetic ground state along the high-symmetry paths in the Brillouin zone of the $R\bar{3}$ conventional cell.}
\label{figband}
\end{figure}

\begin{figure}[bp]
\centerline{\includegraphics[scale=0.17]{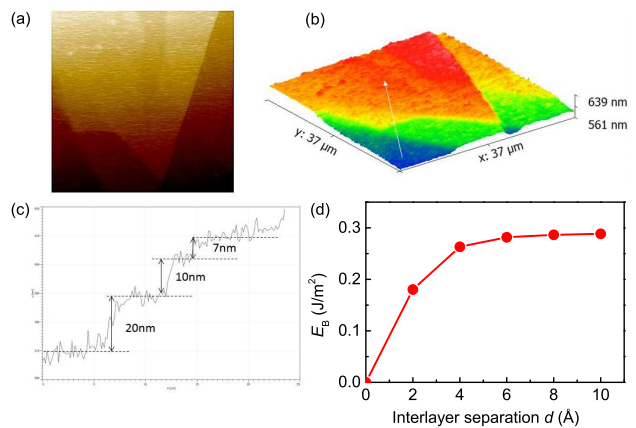}} \vspace*{-0.3cm}
\caption{(a) AFM image of a cleaved thin flake of VI$_{3}$ crystal. (b) Height profile of whole flake. The different colors represent the various heights of surface. (c) Height profile across several step edges along the direction shown by the arrow in (b). (d) Calculated binding energy $E_{\rm B}$ as a function of the interlayer separation $d$.}
\label{flake}
\end{figure}

To further investigate the magnetic and electronic properties of bulk VI$_{3}$ in the $R\bar{3}$ structure at low temperature, we have carried out the spin-polarized first-principles calculations with the experimental lattice constants (Table I) and the relaxed internal atomic positions. By comparing the relative energies among four typical intralayer magnetic configurations (the FM order, the checkerboard AFM N\'eel order, the AFM stripy order, and the AFM zigzag order),\cite{HeJJ} we find that the magnetic ground state of bulk VI$_{3}$ is the FM order, being energetically at least 3.8 meV/V lower than those of the AFM orders. Moreover, our calculations demonstrate that bulk VI$_3$ prefers a weak FM interlayer coupling.
As to the electronic structure, without including the Hubbard $U$, our calculation predicts that VI$_{3}$ should be metallic, different from the insulating behavior observed in experiment. When considering the effect of Hubbard $U$ (3.68 eV),\cite{HeJJ} the calculated band structure of bulk VI$_3$ in the FM ground state shows a direct band gap of $\sim$ 0.84 eV (Fig. 4). This calculation result suggests that bulk VI$_3$ is a FM Mott insulator.

Figure 5(a) displays an atomic force microscopy image of a cleaved thin flake of a VI$_{3}$ crystal and there are several step edges observed on the flake. As shown in Fig. 5(b), the total thickness of this flake is about tens of nanometers and the area along the $ab$ plane is rather large ($\sim$ 40$\times$40 $\mu$m$^{2}$). The hight profile across an edge measured along the white arrow in Fig. 5(b) consists of three steps with heights of 20, 10 and 7 nm (Fig. 5(c)). The minimal step height ($\sim$ 7 nm) is about 10 unit cells of monoclinic structure of VI$_3$ ($h=c\sin (180^\circ -\beta)=$ 6.5888 \AA). It suggests that VI$_{3}$ is easy to be exfoliated down to few layers.
The binding energy $E_{\rm B}$ was calculated by increasing the interlayer distance $d$ of V$_3$ in a $R\bar{3}$ conventional cell.\cite{Eb} The calculated $E_{\rm B}$ as a function of interlayer separation $d$ is shown in Fig. 5(d). The $E_{\rm B}$ increases quickly with increasing $d$ and then converges to the saturation value 0.29 J/m$^2$ (18 meV/\AA$^2$) when $d$ is lager than 6 \AA. This value is in good agreement with previous studies on most vdW crystals ($E_{\rm B}=$ 13 - 21 meV/\AA$^2$).\cite{Eb} Thus it confirms the weak interlayer interaction in VI$_3$, i.e., its easy exfoliation.

\section{Conclusion}

In summary, we synthesized VI$_{3}$ single crystals successfully and studied the structural and physical properties of novel vdW magnetic compound VI$_{3}$ in detail. The structure of VI$_{3}$ at high temperature is monoclinic and changes to rhombohedral at $T_{s}$ = 79 K. The long-range FM order appears at $T_{c}$ = 50 K with the saturate moment close to 2 $\mu_{\rm B}$/V. It suggests that the V$^{3+}$ ion in VI$_{3}$ is at high-spin state with $S=$ 1. Theoretical calculation implies that VI$_{3}$ might be a FM Mott insulator. Moreover, it is easy to be cleaved down to few layers, thus it is a promising candidate of 2D ferromagnet with $S=$ 1. Exploration of magnetic vdW materials with different magnetic moments will not only expand the family of 2D magnets but also deepen the understanding of the relationship between dimensionality and degrees of freedom for spin space.

\section{Acknowledgements}

This work was supported by the National Key R\&D Program of China (Grants No. 2015CB921400, 2016YFA0300504 and 2017YFA0302903), the National Natural Science Foundation of China (Grants No. 11574394, 11774423, 11774424 and 11822412), the Fundamental Research Funds for the Central Universities, and the Research Funds of Renmin University of China (Grants No. 15XNLF06, 15XNLQ07 and 18XNLG14). Computational resources were provided by the Physical Laboratory of High Performance Computing at Renmin University of China.

$^{\dag}$ These authors contributed equally to this work.

$\ast$ zhangxiaobupt@bupt.edu.cn, kliu@ruc.edu.cn, hlei@ruc.edu.cn

\end{document}